\documentclass[prl,twocolumn,floats,aps,epsfig,nofootinbib,amssymb]{revtex4}
\usepackage{subfigure}
\usepackage{graphicx}
\usepackage{cancel}
\usepackage{amssymb}
\usepackage{textcomp}
\usepackage{amsmath}
\usepackage{bm}
\usepackage{times}
\usepackage{epsfig}
\usepackage{color}

\pdfoutput=1
\begin{document}
\title{\Large SUSY Spectrum and the Higgs Mass in the BLMSSM \\ }
\author{\large Pavel Fileviez P{\'e}rez}
\address{Center for Cosmology and Particle Physics (CCPP) \\ 
New York University, 4 Washington Place, NY 10003, USA}
\date{\today}
%
\begin{abstract}
The predictions for the mass of the light CP-even Higgs are investigated in the context of a simple extension 
of the Minimal Supersymmetric Standard Model where the baryon and lepton numbers are local 
gauge symmetries. This theory predicts the existence of light charged and neutral leptons which 
give extra contributions to the Higgs mass at the one-loop level. We show the possibility to satisfy 
the LEP2 bound and achieve a Higgs mass around 125 GeV in a supersymmetric 
spectrum with light sfermions and small left-right mixing in the stop sector. We make a brief discussion of the unique 
leptonic signals at the Large Hadron Collider. This theory predicts baryon number violation at the low scale 
and one could avoid the current LHC bounds on the supersymmetric mass spectrum.
\end{abstract}
\maketitle
\section{I. Introduction}
In the LHC Era there is hope to discover or rule out the existence of a light Higgs boson responsible 
for the spontaneous breaking of the electroweak symmetry in the Standard Model of Particle 
Physics. Recently, the ATLAS and CMS collaborations have discussed possible evidences 
of a Higgs boson with mass around 125 GeV~\cite{ATLAS,CMS}. If these results are 
confirmed one should understand the possibility to predict the properties of the Higgs 
in different theories in order to explain the experimental results.

Supersymmetry (SUSY) is one of the most appealing ideas for physics beyond the 
Standard Model which provides the possibility to understand why the Higgs boson is light. 
In the context of the Minimal Supersymmetric Standard Model (MSSM) the Higgs mass is predicted 
at tree level, $M_h^{tree} = M_Z |\cos 2 \beta |$, but at one-loop level the prediction 
depends mainly on the masses and couplings between the Higgs and the third 
generation of fermions and sfermions. For excellent reviews on Higgs 
phenomenology in SUSY see Refs.~\cite{Djouadi,Carena} and~\cite{Hollik}.
It is well-known that in order to satisfy the LEP2 bound, $M_h > 114.4$ GeV~\cite{LEP2Higgs}, 
the masses of the stops have to be large and/or one must assume a very large left-right mixing in the stop sector. The situation 
is worse if one thinks about the possibility to obtain a Higgs mass around 
125 GeV, as it has been discussed by the ATLAS and CMS collaborations
~\cite{ATLAS,CMS}. Therefore, if one hopes to discover SUSY it is important 
to understand the scenarios where we can satisfy the collider bounds on 
the Higgs mass without assuming a very heavy supersymmetric spectrum.
The new results from ATLAS and CMS have motivated several studies 
in the context of low-energy supersymmetry~\cite{paper1,paper2,paper3,paper4,paper5,paper6,paper7,paper8,paper9,paper10,paper11,paper12,paper13,paper14,paper15,paper16,paper17,paper18}.       

In this letter we investigate the possible predictions for the light CP-even 
Higgs mass and the constraints coming from the ATLAS and CMS results in the 
context of a simple extension of the MSSM, where the baryon and lepton numbers 
are local gauge symmetries spontaneously broken at the TeV scale. We refer to 
this theory as the ``BLMSSM"~\cite{BLMSSM}. In this theoretical framework we can understand 
the absence of the dangerous dimension four and five baryon number violating operators~\cite{review} present in the MSSM. 
One uses a fourth generation that is vector-like with respect to the strong, weak 
and electromagnetic interactions to cancel anomalies and there is no need for large 
Yukawa couplings to be consistent with the experimental limits on fourth generation 
quark masses. Therefore, the model is free of coupling constants with 
Landau poles near the weak scale. Furthermore, there is a natural suppression of flavour 
violation in the quark and leptonic sectors since the gauge symmetries and particle content 
forbid tree level flavor changing neutral currents involving the quarks or charged leptons.  

We find that the existence of new charged and neutral leptons provides the possibility 
to satisfy the LEP2 bound and achieve a Higgs mass around 125 GeV without 
assuming a very heavy spectrum for the supersymmetric particles even in the 
case where the left-right mixing in the stop sector is very small. In this case one has 
new one-loop leptonic contributions for the Higgs mass which allows to increase 
the mass with light stops and we find an upper bound on the ratio between 
the vacuum expectation values of the MSSM Higgses. We discuss the possible 
implications of these results  for the experimental test of supersymmetry 
and the unique leptonic signals at the LHC. 

The letter is organized as follows: In the first section we discuss the main 
features of the BLMSSM, while in the second section we investigate the 
predictions for the Higgs mass and the impact of the new one-loop corrections.
Finally, we summarize the new results.  
%
\section{II. Theoretical Framework}
Recently, we have proposed a simple supersymmetric model where the baryon number 
(B) and lepton number (L) are local gauge symmetries~\cite{BLMSSM}.
This model is based on the gauge symmetry 
$$SU(3)_C \bigotimes SU(2)_L \bigotimes U(1)_Y \bigotimes U(1)_B \bigotimes U(1)_L$$
We refer to this model as the ``BLMSSM". In this context B and L are spontaneously broken gauge symmetries 
at the TeV scale and one can understand the absence of baryon and lepton number violating operators 
of dimension four and five present in the MSSM. 

In this model we have the chiral superfields of the MSSM, and in order to cancel the B and L anomalies we need a vector-like family:  
$\hat{Q}_4$, $\hat{u}_4^c$ ,  $\hat{d}_4^c$, $\hat{L}_4$,  $\hat{e}_4^c$,  $\hat{\nu}^c_4$ and $\hat{Q}_5^c$, $\hat{u}_5$, $\hat{d}_5$, 
$\hat{L}_5^c$, $\hat{e}_5$, $\hat{\nu}_5$. See Table I for the superfields present in the BLMSSM.
It is important to emphasize that one does not need large Yukawa couplings to be consistent with the experimental 
limits on fourth generation quark masses and hence the models are free of coupling constants with Landau poles near the weak scale. 
The baryonic and leptonic anomalies can be cancelled if one imposes the conditions~\cite{BLMSSM}:
\begin{eqnarray}
B_{Q_4}&=& - B_{u^c_4} = - B_{d^c_4}=B_4, \\
B_{Q_5^c} &=& - B_{u_5}= - B_{d_5}, \\ 
B_{Q_4} &+& B_{Q_{5}^c} = -1,
\end{eqnarray} 
in the quark sector and
\begin{eqnarray}
L_{L_4}&=& - L_{e^c_4} = - L_{\nu^c_4}=L_4, \\
L_{L_5^c} &=& - L_{e_5}= - L_{\nu_5}, \\ 
L_{L_4} &+& L_{L_{5}^c} = -3,
\end{eqnarray}
in the leptonic sector. Here $B_i$ and $L_i$ stand for baryon 
and lepton number for a given field, respectively.

\begin{table}[htdp]
\caption{Superfields in the BLMSSM.}
\begin{center}
\begin{tabular}{|c|c|c|c|c|c|}
\hline
Superfields & $SU(3)_C$ & $SU(2)_L$ & $U(1)_Y$ & $U(1)_B$ & $U(1)_L$\\
\hline
\hline
 $\hat{Q}$ & 3 & 2 & 1/6 & 1/3 & 0 \\
 \hline
 $\hat{u}^c$ & $\bar{3}$ & 1 & -2/3 & -1/3 & 0 \\
 \hline
$\hat{d}^c$ & $\bar{3}$ & 1 & 1/3 & -1/3 & 0 \\
\hline
$\hat{L}$ & 1 & 2 & -1/2 & 0 & 1\\
\hline
$\hat{e}^c$ & 1 & 1 & 1 & 0 & -1 \\
\hline
$\hat{\nu}^c$ & 1 & 1 & 0 & 0 & -1 \\
\hline
$\hat{Q}_4$ & 3 & 2 & 1/6 & $B_4$ & 0 \\
\hline
$\hat{u}^c_4$ & $\bar{3}$ & 1 & -2/3 & -$B_4$ & 0 \\
\hline
$\hat{d}^c_4$ & $\bar{3}$ & 1 & 1/3 & -$B_4$ & 0 \\
\hline
$\hat{L}_4$ & 1 & 2 & -1/2 & 0 & $L_4$ \\
\hline
$\hat{e}^c_4$ & 1 & 1 & 1 & 0 & -$L_4$ \\
\hline
$\hat{\nu}^c_4$ & 1 & 1 & 0 & 0 & -$L_4$ \\
\hline
$\hat{Q}_5^c$ & $\bar{3}$ & 2 & -1/6 & -$(1+B_4)$ & 0 \\
\hline
$\hat{u}_5$ & $3$ & 1 & 2/3 &  $1 + B_4$ & 0 \\
\hline
$\hat{d}_5$ & $3$ & 1 & -1/3 & $1 + B_4$ & 0 \\
\hline
$\hat{L}_5^c$ & 1 & 2 & 1/2 & 0 & -$(3 + L_4)$ \\
\hline
$\hat{e}_5$ & 1 & 1 & -1 & 0 & $3 + L_4$ \\
\hline
$\hat{\nu}_5$ & 1 & 1 & 0 & 0 & $3 + L_4$ \\
\hline
$\hat{H}_u$ & 1 & 2 & 1/2 & 0 & 0 \\
\hline
$\hat{H}_d$ & 1 & 2 & -1/2 & 0 & 0 \\
\hline
$\hat{S}_B$ & 1 & 1 & 0 & 1 & 0 \\
\hline
$\hat{\bar{S}}_B$ & 1 & 1 & 0 & -1 & 0 \\
\hline
$\hat{S}_L$ & 1 & 1 & 0 & 0 & -2 \\
\hline
$\hat{\bar{S}}_L$ & 1 & 1 & 0 & 0 & 2 \\
\hline
\end{tabular}
\end{center}
\label{default}
\end{table}%
The full superpotential of the model is given by
\begin{equation}
{\cal W}_{BL}={\cal W}_{MSSM} \ + \  {\cal W}_{Q} \ + \  {\cal W}_{L} \ + \  {\cal W}_{S}, 
\end{equation}
where
\begin{equation}
{\cal W}_{MSSM}=Y_u \hat{Q} \hat{H}_u \hat{u}^c \ + \ Y_d \hat{Q} \hat{H}_d \hat{d}^c \ + \ Y_e \hat{L} \hat{H}_d \hat{e}^c \ + \ \mu \hat{H}_u \hat{H}_d,
\end{equation}
is the MSSM superpotential and
\begin{eqnarray}
{\cal W}_{Q}&=&\lambda_{Q}  \hat{Q}_4  \hat{Q}_5^c \hat{S}_B \ + \  \lambda_{u}  \hat{u}_4^c  \hat{u}_5 \hat{\bar{S}}_B \ + \  \lambda_{d}  \hat{d}_4^c  \hat{d}_5 \hat{\bar{S}}_B,
\end{eqnarray}
is the needed superpotential to give masses to the new quarks in the theory. Notice that we have neglected the interactions of the new quarks with the MSSM Higgses since their 
contributions to the quark masses is not very relevant. In general the Yukawa couplings between the new quarks and the MSSM Higgses can be large and modify the Higgs mass 
at one-loop level\footnote{See Refs.~\cite{Babu} and \cite{Martin} for the study of the Higgs mass in models with vector-like quarks.}. 
Notice that these couplings can have a large impact on the production cross section, $gg \to h$, and it is can be difficult to satisfy the experimental bounds. 
In this letter, we will stick to the most conservative scenario where these quark Yukawa couplings are small and the Higgs mass can be modified only by the Yukawa 
couplings of the new leptons which have to be large in order to satisfy the collider bounds. 

The MSSM Higgs superfields are denoted by $\hat{H}_u \sim (1,2,1/2,0,0)$ and $\hat{H}_d \sim (1,2,-1/2,0,0)$, while 
the new Higgs chiral superfields, $\hat{S}_B \sim (1,1,0,1,0)$ and $\hat{\overline{S}}_B \sim (1,1,0,-1,0)$, generate mass 
for the new leptophobic gauge boson, $Z_B$, and for the new heavy quarks. In the leptonic sector one has the following interactions
\begin{eqnarray}
{\cal W}_{L} &=& Y_{e_{4}}  \  \hat{L}_4 \hat{H}_d \hat{e}^c_4 \ + \  Y_{e_{5}}  \  \hat{L}_5^c \hat{H}_u \hat{e}_5 \ + \ Y_{\nu_{4}} \  \hat{L}_4 \hat{H}_u \hat{\nu}^c_4 
                         \nonumber \\
                         & + & Y_{\nu_{5}}  \  \hat{L}_5^c \hat{H}_d \hat{\nu}_5 
                          +   Y_{\nu}  \  \hat{L}  \hat{H}_u \hat{\nu}^c \ + \  \lambda_{\nu^c}  \  \hat{\nu}^c  \hat{\nu}^c  \hat{\overline{S}}_L.
                          \label{Wlept}
\end{eqnarray}
Here, $\hat{S}_L \sim (1,1,0,0,-2)$ and $\hat{\overline{S}}_L \sim (1,1,0,0,2)$ are chiral superfields which vacuum expectation values 
spontaneously break lepton number and give mass to the quark-phobic gauge boson $Z_L$. 
Notice that we have an implementation of the seesaw mechanism for the light neutrino masses, while the new fourth generation neutrinos have Dirac mass terms.

The mass for the new leptons are given by
\begin{eqnarray}
M_{e_4}&=&Y_{e_4} \frac{v_d}{\sqrt{2}}, \\
M_{e_5} & = & Y_{e_5} \frac{v_u}{\sqrt{2}}, \\
M_{\nu_4}&=&Y_{\nu_4} \frac{v_u}{\sqrt{2}}, 
\end{eqnarray}
and
\begin{eqnarray}
M_{\nu_5} & = & Y_{\nu_5} \frac{v_d}{\sqrt{2}}. 
\end{eqnarray}
Using the above equations and imposing the perturbative condition for the Yukawa coupling, 
$Y_{e_4}^2/4 \pi \leq 1$, one finds an upper bound on $\tan \beta$: $$\tan \beta \leq 6,$$ when $M_{e_4} \geq 100$ GeV. 
This bound on $\tan \beta$ will be crucial to understand the numerical results for the Higgs mass in the next section 
because the upper bound on $\tan \beta$ gives us the largest contribution to the Higgs mass at tree level.

In order to complete our discussion we list the new superpotential in the Higgs sector:
\begin{eqnarray}
{\cal W}_{S} &=& \mu_{L} \hat{\overline{S}}_L \hat{S}_L \ + \   \mu_{B} \hat{\overline{S}}_B \hat{S}_B.
\end{eqnarray}
Now, we could ask the following question: What is the role of the new fermions beyond the anomaly cancellation?
If one takes a look at Eq.(\ref{Wlept}) we can realize that the new charged and neutral leptons can modify the predictions 
for the Higgs masses at one-loop level. As we will show in the next section, their contributions are crucial to 
have light stops and sbottoms in the supersymmetric spectrum.
  
\underline{B and L Symmetry Breaking}: As we have mentioned the symmetries $U(1)_B$ and $U(1)_L$ are broken at the TeV scale since the masses of the new neutral gauge bosons are related  
to the SUSY breaking mass scale. In order to show this we give the dependence of the new gauge boson masses on the  parameters in the model:
\begin{equation}
\frac{1}{2} m_{Z_L}^2 = - |\mu_L|^2 \ + \ \left(  \frac{m_{S_L}^2 \tan^2 \beta_L - m_{\bar{S}_L}^2}{ \tan^2 \beta_L - 1 } \right),
\end{equation} 
and
\begin{equation}
\frac{1}{2} m_{Z_B}^2 = - |\mu_B|^2 \ + \ \left(  \frac{m_{S_B}^2 \tan^2 \beta_B - m_{\bar{S}_B}^2}{ \tan^2 \beta_B - 1 } \right),
\end{equation} 
where $m_{S_L} (m_{S_B})$ and $m_{\bar{S}_L} (m_{\bar{S}_B})$ are soft masses for the Higgses $S_L (S_B)$ and $\bar{S}_L (\bar{S}_B)$, 
while $\tan \beta_L = \left< S_L\right> / \left< \bar{S}_L\right>$ and  $\tan \beta_B = \left< S_B\right> / \left< \bar{S}_B\right>$. 
Note that without the soft SUSY breaking mass terms  one does not get a positive value of $m_{Z_L}^2 (m_{Z_B}^2)$ which implies that  the lepton (baryon) 
number symmetry breaking scale is of order the SUSY breaking scale. 

It is important to mention that in order to break $U(1)_L$ one does not need to introduce the superfields  $\hat{S}_L$ and $\hat{\bar{S}}_L$ since one could assume that 
the right-handed sneutrinos can get a VEV. For the realization of this idea see for example~\cite{SRpV1,SRpV2}. The only problem here is that one could have fast contributions 
to proton decay~\cite{review} if the cutoff of the theory is not very large. Using the equation below, Eq.(14), and the lepton number violating interactions one gets contributions to nucleon decay.
This scenario will be investigated in a future publication.

\underline{Higher-dimensional Operators and Baryon Number Violation}: For any values of the baryonic charges of the new fermions which satisfy the anomaly conditions  the Higgses $\hat{S}_B$ and 
$\hat{\overline{S}}_B$ have charges $1$ and $-1$, respectively. Then, one can write the following dimension five operator which gives rise to baryon number violation
\begin{eqnarray}
\label{bnv}
{\cal W}_B^5 &=& \frac{\tilde{\lambda}^{''}}{\Lambda} \hat{u}^c \hat{d}^c \hat{d}^c \hat{S}_B.
\end{eqnarray}
After $U(1)_B$ breaking at the TeV scale one has the $\lambda^{''}$ MSSM interactions, $\hat{u}^c \hat{d}^c \hat{d}^c$, 
which violate baryon number. As it is well-known, these interactions render the lightest neutralino unstable.  
For example, the photino can decay to a quark and a virtual anti squark that then decays into two quarks. Of course if $\lambda''$ is too small and/or $\Lambda$ is too large this  decay will be not 
occur inside an LHC detector. These interactions can change the current LHC bounds 
on the supersymmetric particles since all constraints on channels with missing energy can be avoided. One interesting scenario is the case where the stop is the LSP and can decay into two jets.

It is important to mention that there is no mixing between the SM fermions and the new fermions from renormalizable operators in the model.  
However, at the dimension five level we could write the following interactions which allow the lightest new quark to decay 
\begin{eqnarray}
{\cal W}_{B}^5 &=& \frac{a_1}{\Lambda} \hat{u}^c_4 \hat{d}^c \hat{d}^c \hat{\overline{S}}_B \ + \   \frac{a_2}{\Lambda} \hat{u}^c \hat{d}^c_4 \hat{d}^c \hat{\overline{S}}_B,
\end{eqnarray}
These interactions require $B_{u^c_4}=5/3$. In this case the new quarks can decay and we can avoid a possible issue with Big Bang Nucleosyntesis.
\section{III. The Light CP-even Higgs Mass}
%
In order to set our notation we define the neutral Higgses as 
\begin{equation}
H_u^0=\frac{1}{\sqrt{2}} \left( v_u \ + \ h_u \right) + \frac{i}{\sqrt{2}} A_u,
\end{equation}
and 
\begin{equation}
H_d^0=\frac{1}{\sqrt{2}} \left( v_d \ + \ h_d \right) + \frac{i}{\sqrt{2} }A_d.
\end{equation}
Using this notation and working in the basis $( h_d, h_u)$ the mass matrix for the MSSM neutral CP-even Higgs is given by
\begin{eqnarray}
&&
{\cal M}_{even}^2= 
\left(
\begin{array}{cc}
	{\cal M}_{11}^2 + \Delta_{11}   
	& 
	{\cal M}_{12}^2 + \Delta_{12}
	\\
	{\cal M}_{12}^2 + \Delta_{12}
	& 
	{\cal M}_{22}^2 + \Delta_{22} 
\end{array}
\right),
\end{eqnarray}
with
\begin{eqnarray}
{\cal M}_{11}^2 &=& M_Z^2 \cos^2  \beta \ + \ M_A^2 \sin^2 \beta, \\
{\cal M}_{12}^2 &=& - (M_A^2 + M_Z^2) \sin \beta \cos \beta, \\
{\cal M}_{22}^2 &=& M_Z^2 \sin^2 \beta \ + \ M_A^2 \cos^2 \beta,
\end{eqnarray}
where $M_A$ is the pseudo-scalar Higgs mass and $\tan \beta = v_u / v_d$. 
In the simplest case when we neglect the left-right mixing in the sfermion sectors, 
$X_i=0$, the one-loop radiative corrections read as
\begin{eqnarray}
\Delta_{11} &=& \frac{3 G_F}{\sqrt{2} \pi^2} \frac{m_b^4}{\cos^2 \beta} \rm{Log} \left( \frac{M_{\tilde{b}_1} M_{\tilde{b}_2}}{m_b^2} \right) 
\nonumber \\
& + &  \frac{G_F}{\sqrt{2} \pi^2} \frac{M^4_{e_4}}{\cos^2 \beta} \rm{Log} \left( \frac{M_{\tilde{e}_4^{1}} M_{\tilde{e}_4^{2}}}{M^2_{e_4}}\right)
\nonumber \\
& + & \frac{G_F}{\sqrt{2} \pi^2} \frac{M^4_{\nu_5}}{\cos^2 \beta} \rm{Log} \left( \frac{M_{\tilde{\nu}_5^{1}} M_{\tilde{\nu}_5^{2}}}{M^2_{\nu_5}}\right),
\end{eqnarray}
and
\begin{eqnarray}
\Delta_{22}&=& \frac{3 G_F}{\sqrt{2} \pi^2} \frac{m_t^4}{\sin^2 \beta} \rm{Log} \left( \frac{M_{\tilde{t}_1} M_{\tilde{t}_2}}{m_t^2} \right) 
\nonumber \\
& + &  \frac{G_F}{\sqrt{2} \pi^2} \frac{M^4_{e_5}}{\sin^2 \beta} \rm{Log} \left( \frac{M_{\tilde{e}_5^{1}} M_{\tilde{e}_5^{2}}}{M^2_{e_5}}\right)
\nonumber \\
& + &  \frac{G_F}{\sqrt{2} \pi^2} \frac{M^4_{\nu_4}}{\sin^2 \beta} \rm{Log} \left( \frac{M_{\tilde{\nu}_4^{1}} M_{\tilde{\nu}_4^{2}}}{M^2_{\nu_4}}\right).
\end{eqnarray}
In our notation $M_{\tilde{b}_i}$  and $M_{\tilde{t}_i}$ are the mass for the sbottoms and stops, 
respectively. We use the same notation for the sfermions in the charged lepton sector. Notice the presence of the 
last terms in the above equations due to new Yukawa couplings between the MSSM Higgses and new charged and 
neutral leptons. In the above matrix we have neglected the contributions proportional to the left-right mixings in the top 
and bottom sectors in order to illustrate the impact of the new contributions. For more general expressions see for example~\cite{H1,H2,H3,H4,H5,H6,H7,H8,H9,H10,H11}.
In our numerical results we will take into account the left-right mixing and we use the most general expressions at one-loop level.

In order to set our notation we define the mass matrix for the sfermions as
\begin{eqnarray}
&&
{\cal M}_{\tilde{f}}^2= 
\left(
\begin{array}{cc}
	m_f^2 + m_{LL}^2   
	& 
	m_f  \ X_f
	\\
	m_f \ X_f
	& 
	m_f^2 + m_{RR}^2
\end{array}
\right),
\end{eqnarray}
where
\begin{eqnarray}
m_{LL}^2 &=& m_{\tilde{f}_L}^2 \ + \ (I_3^f  -  Q_f  \sin^2 \theta_W) M_Z^2 \cos 2 \beta, \\
m_{RR}^2 &=& m_{\tilde{f}_R}^2 \ + \ Q_f \sin^2 \theta_W M_Z^2 \cos 2 \beta, \\
X_f &=& a_f \ + \  \mu (\tan \beta)^{-2 I_3^f}, 
\end{eqnarray}
where $I_3^f$ is the isospin, $a_f$ are the trilinear soft terms, $m_{\tilde{f}_L}$ and $m_{\tilde{f}_R}$ are the soft masses and 
the mass for the physical states are given by
\begin{eqnarray}
m_{\tilde{f}_{1,2}}^2 &=& m_f^2 \ + \ \frac{1}{2} \left(  m_{LL}^2 +  m_{RR}^2 \right)  \nonumber \\
& \mp& \frac{1}{2}   \sqrt{(m_{LL}^2 - m_{RR}^2)^2 + 4 m_f^2 X_f^2}.
\end{eqnarray}
Notice that the contributions from the $U(1)_B$ and $U(1)_L$ D-terms to the sfermion masses can be absorbed in the soft masses.
Here we stick to this possibility for simplicity.
Now, it is easy to realize that the existence of new leptons in the theory will help to increase the mass for the light Higgs in the 
model and allows for a light stop and sbottom in the spectrum. We have investigated this issue in detail and show our main results.    

\begin{itemize}

\item Minimal Mixing Scenario: $X_i=0$

In Fig.1 we show the predictions for the mass of the lightest CP-even Higgs for different values of $\tan \beta$ assuming a simple scenario where 
all soft sfermion masses are taken equal to 500 GeV and we have neglected the left-right mixings in the stop, sbottom and heavy new slepton sectors.
In blue we show the results in the context of the MSSM, while the line in black shows the predictions in the BLMSSM.
One can appreciate that it is possible to increase the mass of the Higgs in more than 10 GeV in the majority fraction of the parameter space. 
Therefore, we can say that it is possible to satisfy easily the LEP2 bound, $M_h > 114.4$ GeV, and achieve a mass around 125 GeV due 
to the extra contributions. There results are relevant if one thinks about the possibility to discover a light SUSY spectrum where all squarks 
of the third generation are light enough to be discovered at the LHC and the possibility to explain the existence of a light Higgs boson 
without having a lot of fine-tuning.

In order to understand the possibility to have a Higgs mass around 125 GeV in this model, we show in Fig.2 the 
allowed values for the masses of the new leptons using the same input parameters as in Fig.1. Here $\tan \beta$ changes 
between 4 and 7 and it is important to appreciate that the new leptons can be light with masses approximately 
between 150 GeV and 180 GeV. In this way we show the possibility to have a consistent scenario in agreement 
with all constraints without assuming very heavy stops and sbottoms.  

\begin{figure}[h]
\includegraphics[scale=1,width=9.0cm]{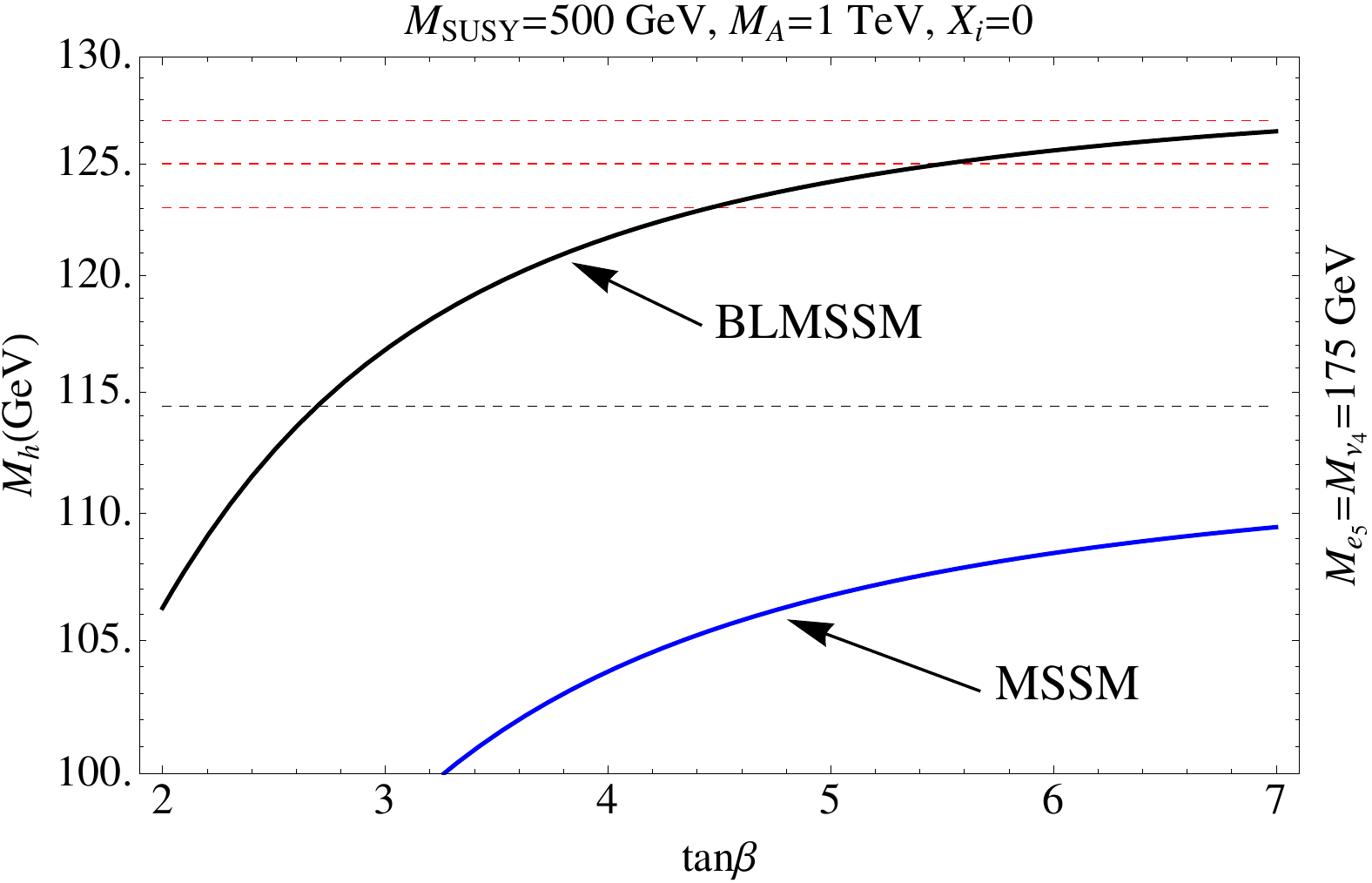}
\caption{Prediction for the light Higgs boson mass for different values of $\tan \beta$. Here it is assumed 
the same soft mass for the third generation of MSSM sfermions and the new sleptons, $M_{\rm{SUSY}}=500$ GeV. 
The mass of the CP-odd Higgs is assumed to be equal to 1 TeV and the left-right mixing in the sfermion 
sector is neglected. The lepton masses are $M_{\nu_5}=M_{e_4}=100$ GeV. We use $M_t=172.9$ GeV as the pole mass for the top quark.}
\end{figure}

\begin{figure}[h]
\includegraphics[scale=1,width=7.5cm]{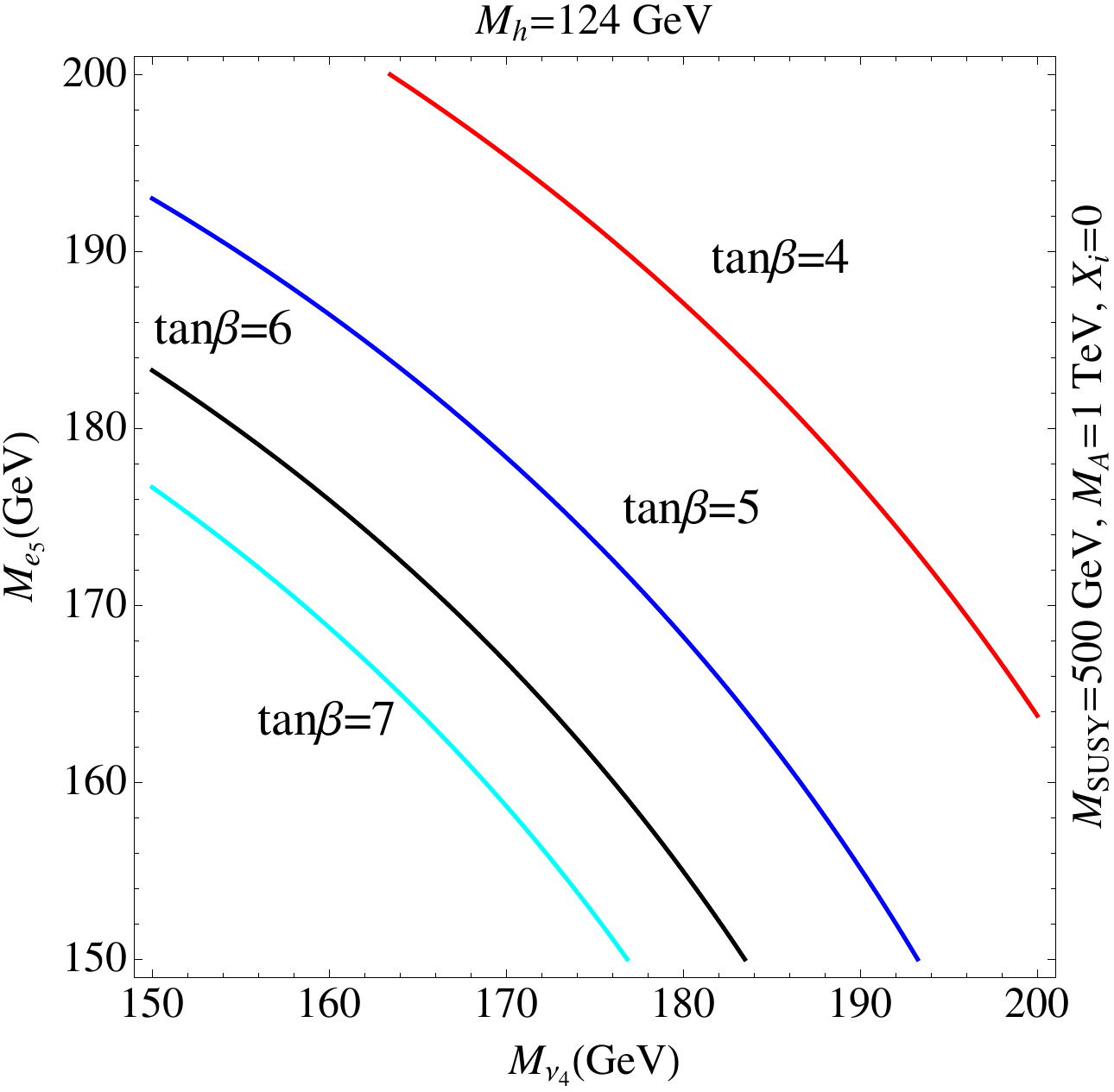}
\caption{Allowed values for the new lepton masses when the Higgs mass is equal to 124 GeV and 
$\tan \beta$ changes between 4 and 7. Here we have assumed the same input parameters as in Fig. 1.}
\label{Fig2}
\end{figure}

\begin{figure}[h]
\includegraphics[scale=1,width=9.0cm]{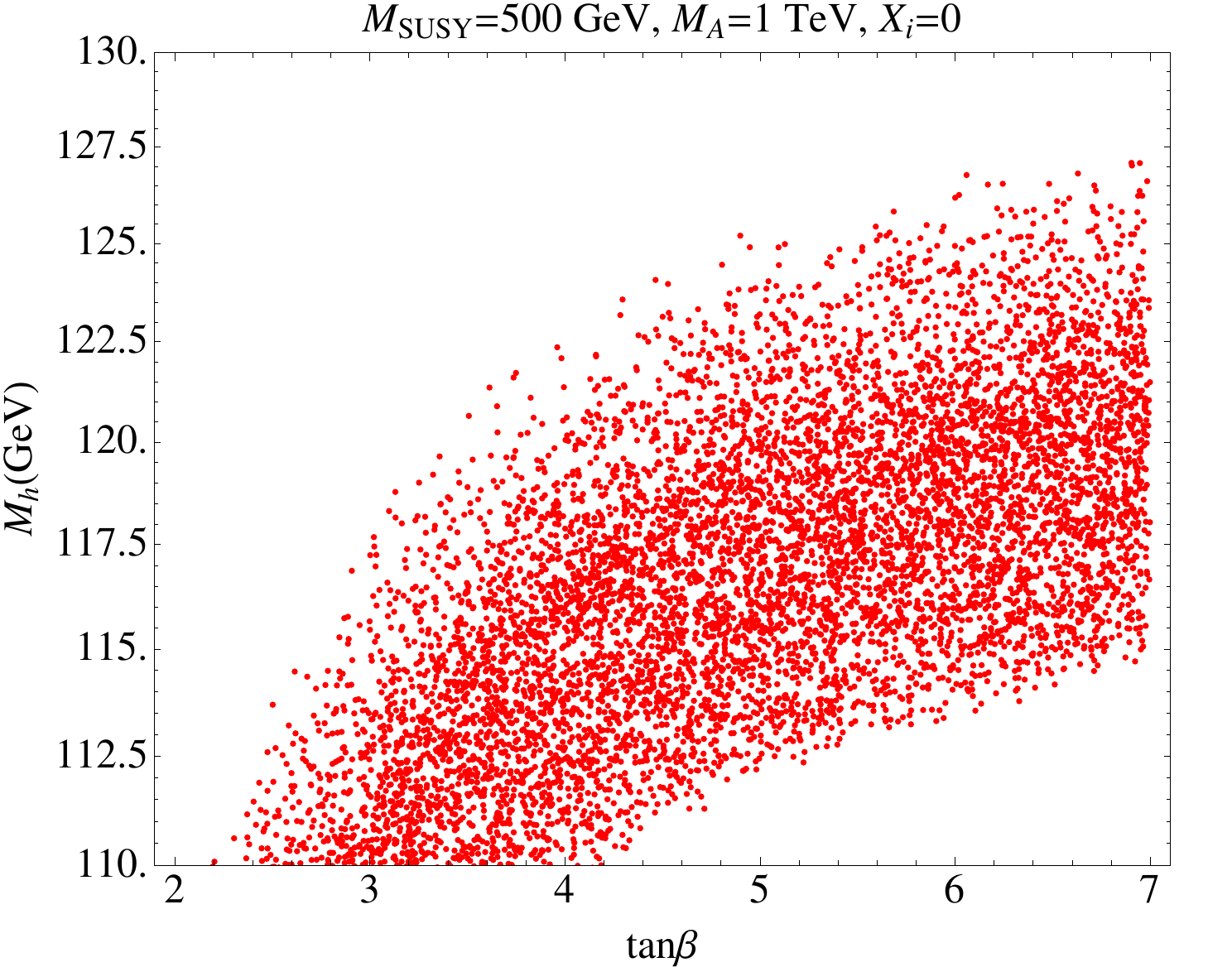}
\caption{The Higgs boson mass versus $\tan \beta$ when $M_{\rm{SUSY}}=500$ GeV, zero left-right mixing and the leptonic masses, $M_{e_5}$ and $M_{\nu_4}$, change between 100 GeV and 180 GeV.}
\label{Fig3}
\end{figure}

Finally, we compute the Higgs mass changing the values of the lepton masses, $M_{e_5}$, and $M_{\nu_4}$, between 100 GeV and 180 GeV.
The numerical results are shown in Fig. 3. One can appreciate that the Higgs mass can be as large as 127 GeV and one can have several 
solutions where the Higgs mass is the range $124 \pm 3$ GeV. It is important to mention that the fourth generation neutrino in general could mix 
with the SM neutrinos and we can avoid a stable charged lepton even in the case when $M_{\nu_4} > M_{e_4}$.

\item `Natural' Mixing Scenario: $X_t= X_b = M_{\rm{SUSY}}$ 

Let us investigate the possible predictions when one takes into account the left-right mixing 
in the stop and sbottom sector. In this case we choose $X_t=X_b=M_{\rm{SUSY}}$ and neglect 
the left-right mixing in the new leptonic sector. In Fig.4 we show the results of the Higgs mass changing 
$\tan \beta$ and assuming very light leptons, $M_{e_5}=M_{\nu_4}=140$ GeV. As in the previous 
scenario, the blue line corresponds to the results in the MSSM and the black line shows the predictions 
in the BLMSSM. Again, one can show that it is possible to get a Higgs mass around 125 GeV with very light 
leptons when the left-right mixing in the stop sector is small. Here we chose 
$X_t = M_{\rm{SUSY}}$, since one could expect that all SUSY breaking terms are of the same size.
Notice that the case of gauge mediation is different since the trilinear terms are typically very small.  
\begin{figure}[h]
\includegraphics[scale=1,width=9.0cm]{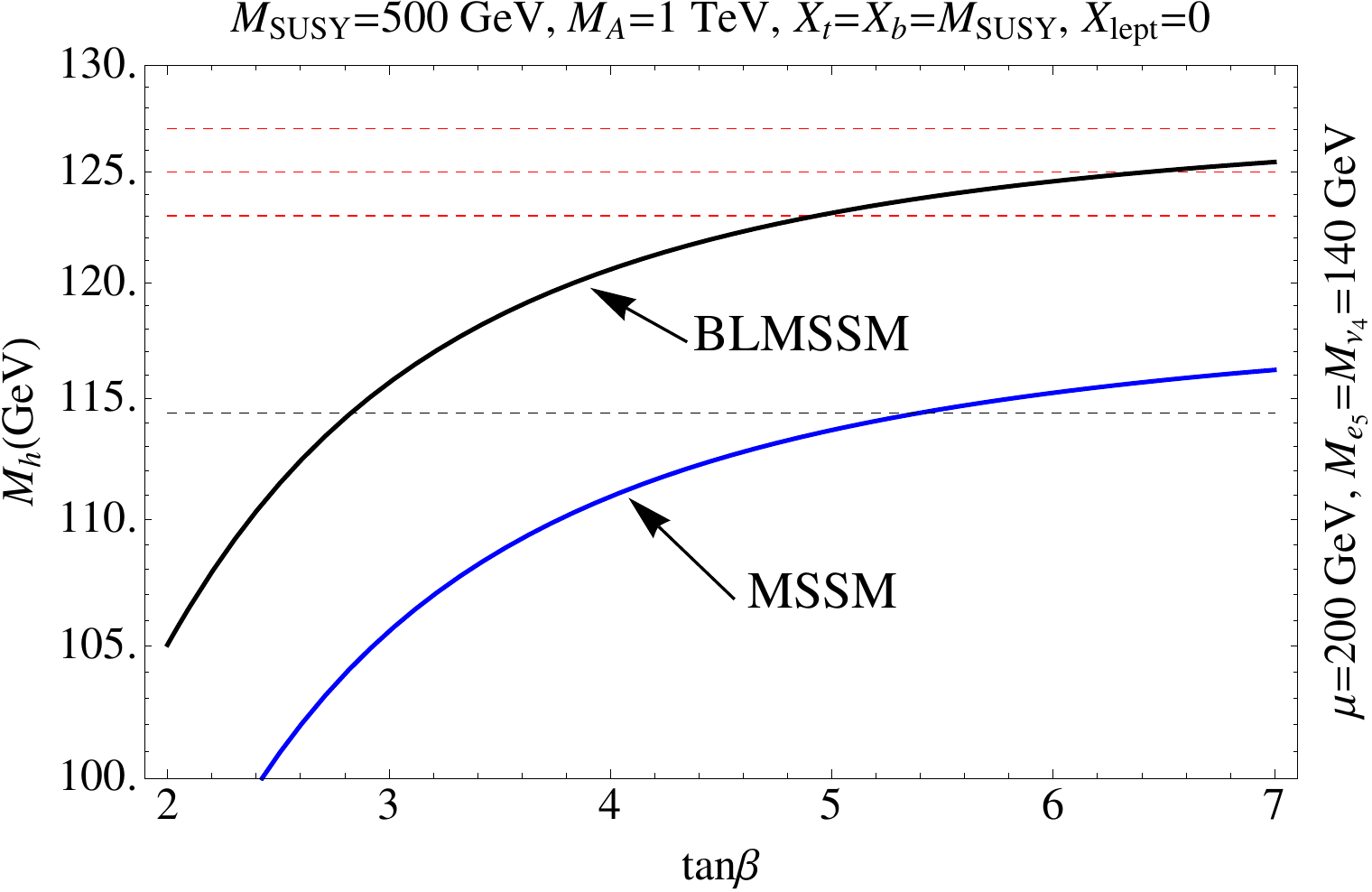}
\caption{The Higgs mass versus $\tan \beta$ where the left-right mixing in the stop and sbottom sector 
is taken equal to the soft sfermion mass $M_{\rm{SUSY}}$. The mass of the leptons are, 
$M_{\nu_4}=M_{e_5}=140$ GeV, and the mixing in the leptonic sector is neglected.}
\end{figure}
In order to complete our discussion we show in Fig.5 the allowed values for the new 
charged and neutral leptons when the Higgs mass is 125 GeV. In this plot $\tan \beta$ 
changes between 3 and 6. It is obvious that in this scenario the new leptons could be 
very light and we can expect the possible discovery at the LHC. In summary, one can 
say that in this scenario the new leptons can be lighter since the left-right 
mixing in the stop sector is not very small. These results are very important if one wants 
to understand the possible evolution of the gauge and Yukawa couplings up to 
the high scale close to the unification scale.   

\begin{figure}[tb]
\includegraphics[scale=1,width=7.5cm]{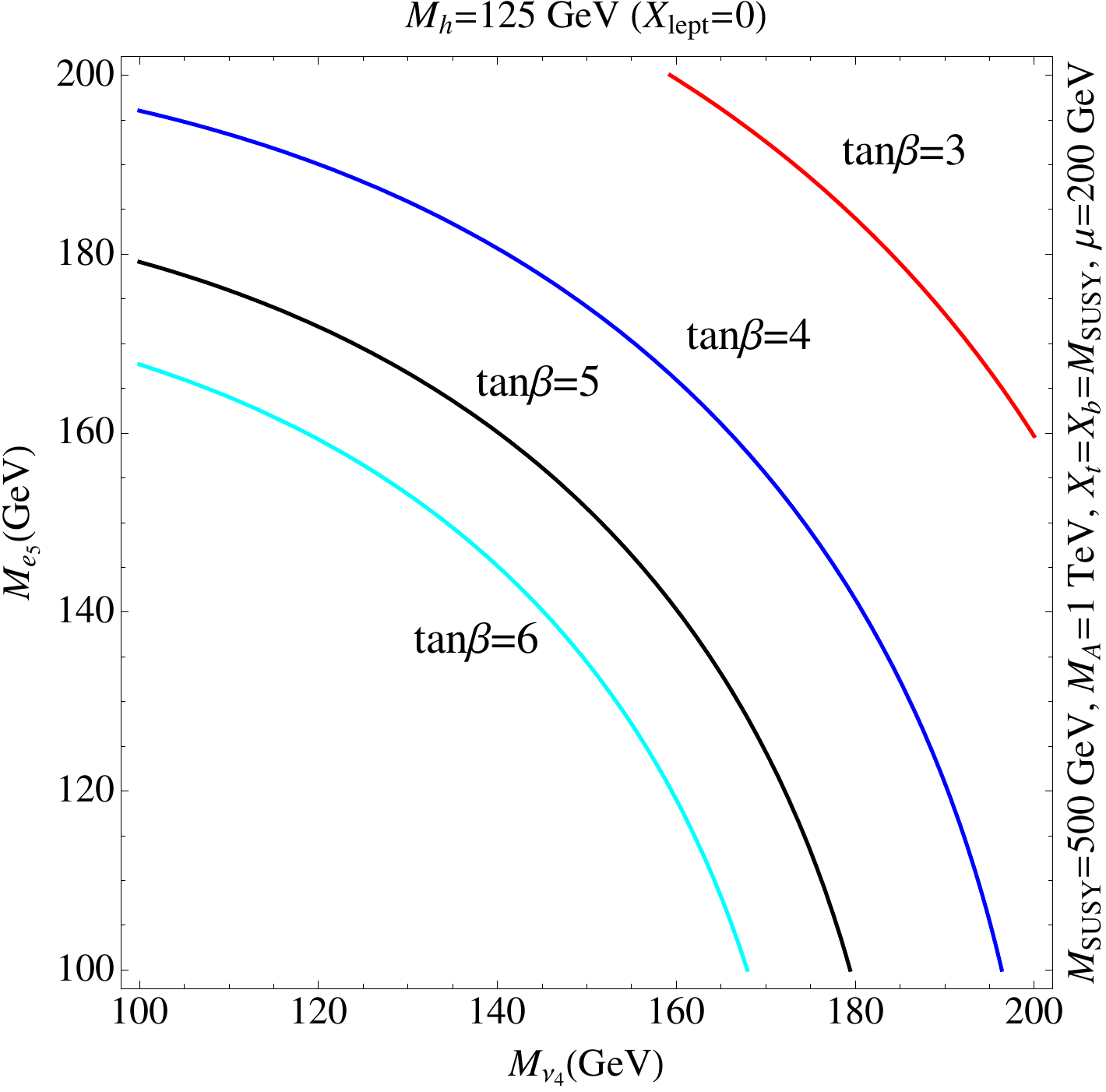}
\caption{Allowed values for the new lepton masses when the Higgs mass is equal to 125 GeV 
when $\tan \beta$ changes between 2 and 6. Here we have assumed the same input parameters listed 
in Fig. 3 and $\mu=200$ GeV.}
\end{figure} 
\begin{figure}[h]
\includegraphics[scale=1,width=9.0cm]{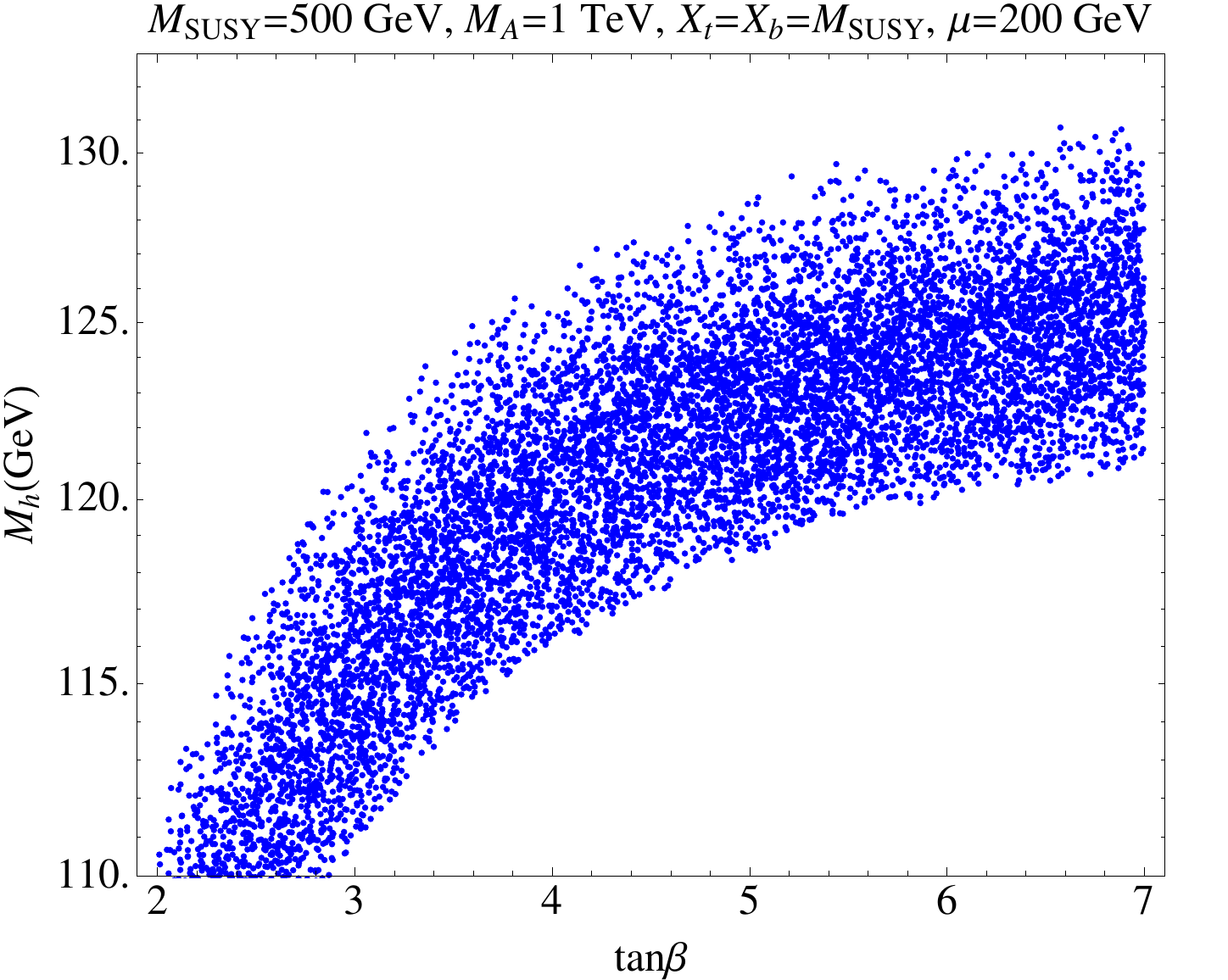}
\caption{The Higgs boson mass versus $\tan \beta$ when $M_{\rm{SUSY}}=500$ GeV, $X_t=X_b=M_{SUSY}$, zero left-right mixing in the leptonic sector 
and all lepton masses change between 100 GeV and 170 GeV.}
\label{Fig6}
\end{figure}
As in the previous scenario, in Fig.6 we show the values for the Higgs mass when the lepton masses change between 100 GeV and 170 GeV.
In this way we can see that the Higgs mass can be much larger, and the upper bound in this case is around 130 GeV. Notice that there are 
many solutions in the range $125 \pm 4$ GeV.  
\item Non-zero left-right sleptonic mixing

Now, let us investigate the predictions for the Higgs mass where one takes into account the left-right 
mixing in the heavy sleptonic sector. In order to illustrate the numerical solutions we choose 
$X_t=X_b=X_{\rm{lept}}=M_{\rm{SUSY}}$. In Fig. 7 we show the results for the Higgs mass for different values 
of tan$\beta$ where the mass of the leptons is $130$ GeV and a negative value 
for the $\mu$ parameter. In this way we can achieve a Higgs mass around 125 GeV for the region when $\tan \beta$ is between 4 and 5.
\begin{figure}[h]
\includegraphics[scale=1,width=9.0cm]{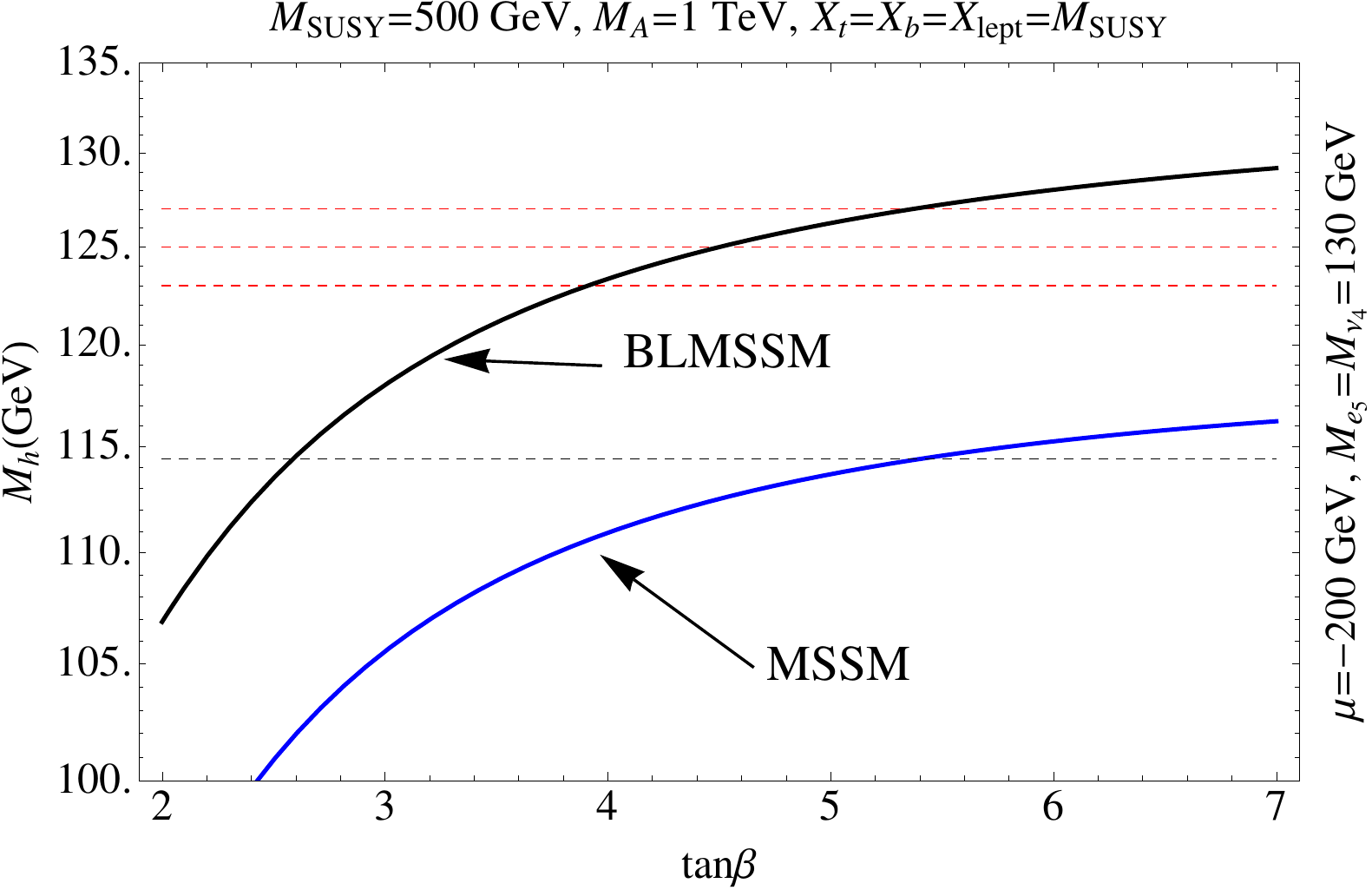}
\caption{The Higgs mass versus $\tan \beta$ where the left-right mixing in the stop, sbottom and 
leptonic sectors are taken equal to $M_{\rm{SUSY}}$, and $\mu=-200$ GeV.}
\end{figure} 
\begin{figure}[h]
\includegraphics[scale=1,width=9.0cm]{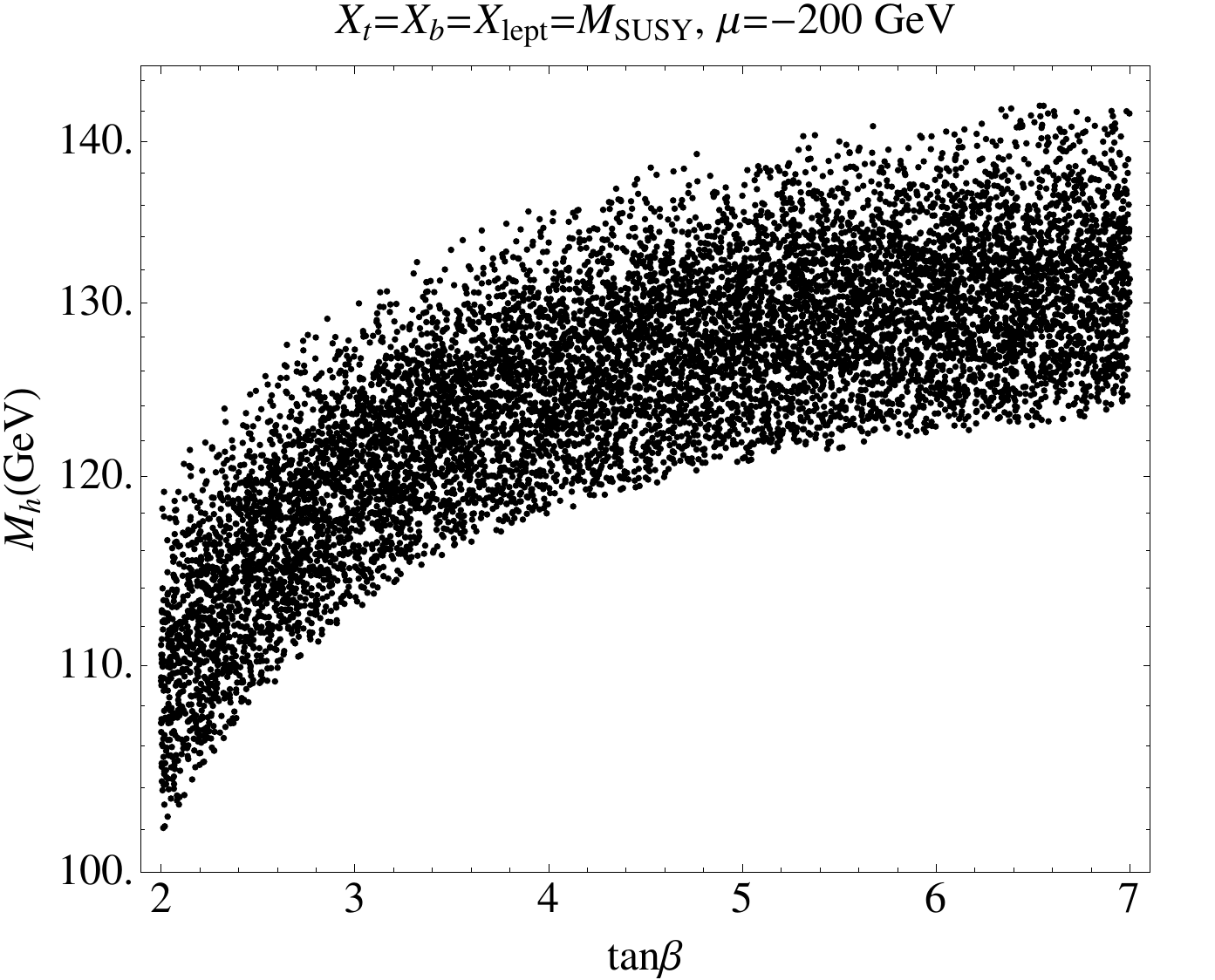}
\caption{The Higgs boson mass versus $\tan \beta$ when $M_{\rm{SUSY}}=500$ GeV, $X_t=X_b=X_{\rm{lept}}=M_{\rm{SUSY}}$, 
and the lepton masses, $M_{\nu_4}$ and $M_{e_5}$, change between 100 GeV and 170 GeV.}
\label{Fig6}
\end{figure}
In Fig. 8 we show the numerical results for the Higgs mass when all lepton masses change between 100 GeV and 170 GeV.
In this case the Higgs mass can be much larger in the region where $\tan \beta$ changes between 2 and 7. 
The upper bound on the Higgs mass in this case is around 140 GeV without assuming very heavy squarks.  
\end{itemize}
These results tell us that due to the existence of the new light charged and neutral leptons, which are needed to 
cancel the leptonic anomalies, the Higgs mass can be much larger than in the MSSM and one can satisfy 
the experimental bounds without assuming very heavy stops and sbottoms.  
It is important to mention that in our previous paper~\cite{BLMSSM} we have investigated the constraints 
coming from the electroweak precision constraints, the S, T and U parameters, and we have showed that the new leptons can be light in agreement 
with the experimental constraints.
\\
\\
{\bf{Heavy Leptons at the LHC}}.
It is important to understand how to test the existence of the leptons predicted by the theory.
The bounds on the heavy lepton mass are basically, $M > 100$ GeV, see for example~\cite{LEP2}.
In our case the bounds should be different because the heavy Dirac neutrinos are stable and the charged 
leptons do not decay into the SM leptons. As we have mentioned, the new Dirac neutrinos are stable 
since the extra generation of leptons never mix with the SM leptons. Therefore, one expects different 
channels with missing energy. Here we list three different channels:
\begin{itemize}
\item Four jets and Missing Energy: The charged leptons, $e_4$, can be produced in pair through the photon, the $Z$ or the Higgs boson.
Then, one can have the channels
$$q \bar{q} \ \to \  \gamma, Z \  \to  \bar{e}_4 e_4 \  \to \  W^+ W^- \bar{\nu}_4 \nu_4 \  \to \  4j \bar{\nu}_4 \nu_4, $$
$$ gg \  \to \  h \  \to  \bar{e}_4 e_4 \  \to \  W^+ W^- \bar{\nu}_4 \nu_4 \  \to \  4j \bar{\nu}_4 \nu_4. $$
\item Two Jets plus Missing Energy: One can have the associated production of a charged lepton and neutrino through the $W$ 
gauge boson 
$$ \bar{q} q^{'} \ \to \ W \ \to \bar{e}_4 \nu_4 \  \to \  W^+  \bar{\nu}_4 \nu_4 \  \to \  2j  \bar{\nu}_4 \nu_4 .$$
\item Channels with Missing Energy:
One can produce the heavy stable neutrinos through the $Z$ or $h$ boson
$$q \bar{q} \ \to \ Z \  \to  \bar{\nu}_4 \nu_4, \  \  gg \  \to \  h \  \to \bar{\nu}_4 \nu_4. $$
In this case one should use initial state radiation (ISR), a photon or a jet, in order to look for these events with missing energy.
\end{itemize}
Notice one can get the same signals for the leptons of the other generation. The cross sections above are much smaller than the cross section for squarks production, 
but one should be able to use the searches for multijets and missing energy to set bounds on the heavy lepton masses. 
A detailed analysis of these channels is beyond the scope of this letter. For a study of the possible signals of heavy 
leptons in SUSY models with fourth generations see Ref.~\cite{Nath}.

It is important to mention that the existence of light charged leptons and the selectrons affect the predictions for the Higgs decays, 
$h \to \gamma \gamma$~\cite{Mark}, and  $h \to Z \gamma$. It is not difficult to show that the branching ratio for the Higgs into 
gamma gamma decay is not very different from the prediction in the Standard Model. This issue will be investigated in a future publication. 
%
\section{IV. Summary and Outlook}
We have studied the predictions for the mass of the light neutral CP-even Higgs 
in the context of a simple extension of the Minimal Supersymmetric Standard Model 
where the baryon and lepton numbers are local gauge symmetries. We refer to this 
theory as the BLMSSM. This theory predicts the existence of light charged and neutral leptons 
which give extra contributions to the Higgs mass at the one-loop level. One finds an upper bound 
on $\tan \beta$ coming from the perturbativity condition on the Yukawa couplings for the heavy leptons.
We have shown the possibility to satisfy the LEP2 bound and/or achieve a Higgs mass around 125 GeV in a supersymmetric spectrum 
with light sfermions and small left-mixing in the stop sector. These results are in agreement with the electroweak precision constraints and collider experiments.
This theory predicts baryon number violation at the low scale and one can avoid the current LHC bounds on the supersymmetric mass spectrum 
based on the searches for channels with multijets and missing energy. 

The light charged leptons and their superpartners affect the predictions for the Higgs decays into two photons 
and a $Z$ and a photon. One should investigate these decays in great detail in order to make possible predictions 
for the LHC experiments. In this letter we did not investigate the behavior of the Yukawa couplings at the 
high scale, but since we are interested in the scenarios where there is not a Landau pole at the low scale 
we have shown our numerical results assuming very light charged leptons. These aspects of the theory will be 
investigated in a future publication. In this work we did not stick to any specific SUSY breaking scenario, 
such as MSUGRA, this analysis is beyond the scope of this letter.

{\textit{Acknowledgments}}:
{\small I would like to thank Pran Nath, Sogee Spinner, Neal Weiner and Mark B. Wise for discussions. 
This work has been supported by the James Arthur Fellowship, CCPP, New York University.}


\begin{thebibliography}{000}

\bibitem{ATLAS} F. Gianotti, CERN Public Seminar, 
``Update on the Standard Model Higgs searches in ATLAS," December 2011.

\bibitem{CMS} G. Tonelli, CERN Public Seminar, 
``Update on the Standard Model Higgs searches in CMS," December 2011.

\bibitem{Djouadi}
  A.~Djouadi,
  ``The Anatomy of electro-weak symmetry breaking. II. The Higgs bosons in the minimal supersymmetric model,''
  Phys.\ Rept.\  {\bf 459} (2008) 1
  [hep-ph/0503173].

\bibitem{Carena}
  M.~S.~Carena and H.~E.~Haber,
  ``Higgs boson theory and phenomenology,''
  Prog.\ Part.\ Nucl.\ Phys.\  {\bf 50} (2003) 63
  [hep-ph/0208209].

\bibitem{Hollik}
  S.~Heinemeyer, W.~Hollik and G.~Weiglein,
  ``Electroweak precision observables in the minimal supersymmetric standard model,''
  Phys.\ Rept.\  {\bf 425} (2006) 265
  [hep-ph/0412214].
  
\bibitem{LEP2Higgs}
  R.~Barate {\it et al.}  [LEP Working Group for Higgs boson searches and ALEPH and DELPHI and L3 and OPAL Collaborations],
  ``Search for the standard model Higgs boson at LEP,''
  Phys.\ Lett.\ B {\bf 565} (2003) 61
  [hep-ex/0306033].
  
  
\bibitem{paper1}
  S.~Heinemeyer, O.~Stal and G.~Weiglein,
  ``Interpreting the LHC Higgs Search Results in the MSSM,''
  arXiv:1112.3026 [hep-ph].
     
\bibitem{paper2}
  M.~Ibe and T.~T.~Yanagida,
  ``The Lightest Higgs Boson Mass in Pure Gravity Mediation Model,''
  arXiv:1112.2462 [hep-ph].
  
\bibitem{paper3}
  A.~Arbey, M.~Battaglia, A.~Djouadi, F.~Mahmoudi and J.~Quevillon,
  ``Implications of a 125 GeV Higgs for supersymmetric models,''
  arXiv:1112.3028 [hep-ph].

\bibitem{paper4}
  J.~L.~Feng, K.~T.~Matchev and D.~Sanford,
  ``Focus Point Supersymmetry Redux,''
  arXiv:1112.3021 [hep-ph].

\bibitem{paper5}
  H.~Baer, V.~Barger and A.~Mustafayev,
  ``Implications of a 125 GeV Higgs scalar for LHC SUSY and neutralino dark
  matter searches,''
  arXiv:1112.3017 [hep-ph].

\bibitem{paper6}
  M.~Carena, S.~Gori, N.~R.~Shah and C.~E.~M.~Wagner,
  ``A 125 GeV SM-like Higgs in the MSSM and the $\gamma \gamma$ rate,''
  arXiv:1112.3336 [hep-ph].

\bibitem{paper7}
  P.~Draper, P.~Meade, M.~Reece and D.~Shih,
  ``Implications of a 125 GeV Higgs for the MSSM and Low-Scale SUSY Breaking,''
  arXiv:1112.3068 [hep-ph].

\bibitem{paper8}
  L.~J.~Hall, D.~Pinner and J.~T.~Ruderman,
  ``A Natural SUSY Higgs Near 126 GeV,''
  arXiv:1112.2703 [hep-ph].

\bibitem{paper9}
  G.~Kane, P.~Kumar, R.~Lu and B.~Zheng,
  ``Higgs Mass Prediction for Realistic String/M Theory Vacua,''
  arXiv:1112.1059 [hep-ph].

\bibitem{paper10}
  S.~Akula, B.~Altunkaynak, D.~Feldman, P.~Nath and G.~Peim,
  ``Higgs Boson Mass Predictions in SUGRA Unification, Recent LHC-7 Results,
  and Dark Matter,''
  arXiv:1112.3645 [hep-ph].

\bibitem{paper11}
  O.~Buchmueller {\it et al.},
  ``Higgs and Supersymmetry,''
  arXiv:1112.3564 [hep-ph].
 
\bibitem{paper12}
  U.~Ellwanger,
  ``A Higgs boson near 125 GeV with enhanced di-photon signal in the NMSSM,''
  arXiv:1112.3548 [hep-ph].
  
\bibitem{paper13}
  G.~F.~Giudice and A.~Strumia,
  ``Probing High-Scale and Split Supersymmetry with Higgs Mass Measurements,''
  arXiv:1108.6077 [hep-ph].
  
 \bibitem{paper14}
 I.~Gogoladze, Q.~Shafi and C.~S.~Un,
 ``Higgs Boson Mass from t-b-tau Yukawa Unification,''
 arXiv:1112.2206 [hep-ph].
 
\bibitem{paper15}
  A.~Arbey, M.~Battaglia and F.~Mahmoudi,
  ``Constraints on the MSSM from the Higgs Sector - A pMSSM Study of Higgs Searches, Bs to mu+ mu- and Dark Matter Direct Detection,''
  arXiv:1112.3032 [hep-ph].
  
\bibitem{paper16}
  M.~Kadastik, K.~Kannike, A.~Racioppi and M.~Raidal,
  ``Implications of the 125 GeV Higgs boson for scalar dark matter and for the CMSSM phenomenology,''
  arXiv:1112.3647 [hep-ph].
  
\bibitem{paper17}
  J.~Cao, Z.~Heng, D.~Li and J.~M.~Yang,
  ``Current experimental constraints on the lightest Higgs boson mass in the constrained MSSM,''
  arXiv:1112.4391 [hep-ph].
  
\bibitem{paper18}
  A.~Arvanitaki and G.~Villadoro,
  ``A Non Standard Model Higgs at the LHC as a Sign of Naturalness,''
  arXiv:1112.4835 [hep-ph].
    
     
\bibitem{BLMSSM} 
  P.~Fileviez Perez and M.~B.~Wise,
  ``Breaking Local Baryon and Lepton Number at the TeV Scale,''
  JHEP {\bf 1108}, 068 (2011)
  [arXiv:1106.0343 [hep-ph]].
  
\bibitem{review}
  P.~Nath and P.~Fileviez P\'erez,
  ``Proton stability in grand unified theories, in strings, and in branes,''
  Phys.\ Rept.\  {\bf 441} (2007) 191;
 [arXiv:hep-ph/0601023].
  
\bibitem{Babu}
  K.~S.~Babu, I.~Gogoladze, M.~U.~Rehman and Q.~Shafi,
  ``Higgs Boson Mass, Sparticle Spectrum and Little Hierarchy Problem in Extended MSSM,''
  Phys.\ Rev.\ D {\bf 78} (2008) 055017
  [arXiv:0807.3055 [hep-ph]].
  
\bibitem{Martin}
  S.~P.~Martin,
  ``Extra vector-like matter and the lightest Higgs scalar boson mass in low-energy supersymmetry,''
  Phys.\ Rev.\ D {\bf 81} (2010) 035004
  [arXiv:0910.2732 [hep-ph]].
    
\bibitem{SRpV1}
  P.~Fileviez Perez and S.~Spinner,
  ``Spontaneous R-Parity Breaking and Left-Right Symmetry,''
  Phys.\ Lett.\ B {\bf 673} (2009) 251
  [arXiv:0811.3424 [hep-ph]].

\bibitem{SRpV2}
  V.~Barger, P.~Fileviez Perez and S.~Spinner,
  ``Minimal gauged U(1)(B-L) model with spontaneous R-parity violation,''
  Phys.\ Rev.\ Lett.\  {\bf 102} (2009) 181802
  [arXiv:0812.3661 [hep-ph]].

\bibitem{H1}
  Y.~Okada, M.~Yamaguchi and T.~Yanagida,
  ``Upper bound of the lightest Higgs boson mass in the minimal supersymmetric standard model,''
  Prog.\ Theor.\ Phys.\  {\bf 85} (1991) 1.

\bibitem{H2}  
  J.~R.~Ellis, G.~Ridolfi and F.~Zwirner,
  ``Radiative corrections to the masses of supersymmetric Higgs bosons,''
  Phys.\ Lett.\ B {\bf 257} (1991) 83.
  
\bibitem{H3}    
  S.~P.~Li and M.~Sher,
  ``Upper Limit To The Lightest Higgs Mass In Supersymmetric Models,''
  Phys.\ Lett.\ B {\bf 140} (1984) 33.
  
\bibitem{H4}   
  R.~Barbieri and M.~Frigeni,
  ``The Supersymmetric Higgs searches at LEP after radiative corrections,''
  Phys.\ Lett.\ B {\bf 258} (1991) 395.
  
\bibitem{H5}    
  M.~Drees and M.~M.~Nojiri,
  ``One loop corrections to the Higgs sector in minimal supergravity models,''
  Phys.\ Rev.\ D {\bf 45} (1992) 2482.
  
\bibitem{H6}   
J.~A.~Casas, J.~R.~Espinosa, M.~Quiros and A.~Riotto,
  ``The Lightest Higgs boson mass in the minimal supersymmetric standard model,''
  Nucl.\ Phys.\ B {\bf 436} (1995) 3
   [Erratum-ibid.\ B {\bf 439} (1995) 466]
  [hep-ph/9407389].
  
\bibitem{H7}   
J.~R.~Ellis, G.~Ridolfi and F.~Zwirner,
  ``On radiative corrections to supersymmetric Higgs boson masses and their implications for LEP searches,''
  Phys.\ Lett.\ B {\bf 262} (1991) 477.
  
\bibitem{H8}
  M.~A.~Diaz and H.~E.~Haber,
  ``Can the Higgs mass be entirely due to radiative corrections?,''
  Phys.\ Rev.\ D {\bf 46} (1992) 3086.
  
\bibitem{H9}
  M.~S.~Carena, M.~Quiros and C.~E.~M.~Wagner,
  ``Effective potential methods and the Higgs mass spectrum in the MSSM,''
  Nucl.\ Phys.\ B {\bf 461} (1996) 407
  [hep-ph/9508343].
  
\bibitem{H10}
  S.~Heinemeyer, W.~Hollik and G.~Weiglein,
  ``The Masses of the neutral CP - even Higgs bosons in the MSSM: Accurate analysis at the two loop level,''
  Eur.\ Phys.\ J.\ C {\bf 9} (1999) 343
  [hep-ph/9812472].
  
\bibitem{H11}
  G.~Degrassi, S.~Heinemeyer, W.~Hollik, P.~Slavich and G.~Weiglein,
  ``Towards high precision predictions for the MSSM Higgs sector,''
  Eur.\ Phys.\ J.\ C {\bf 28} (2003) 133
  [hep-ph/0212020].
  
 
\bibitem{LEP2}
  P.~Achard {\it et al.}  [L3 Collaboration],
  ``Search for heavy neutral and charged leptons in $e^{+} e^{-}$ annihilation at LEP,''
  Phys.\ Lett.\ B {\bf 517} (2001) 75
  [hep-ex/0107015].
    
\bibitem{Nath}
  T.~Ibrahim and P.~Nath,
  ``An MSSM Extension with a Mirror Fourth Generation, Neutrino Magnetic Moments and LHC Signatures,''
  Phys.\ Rev.\ D {\bf 78} (2008) 075013
  [arXiv:0806.3880 [hep-ph]].
  
\bibitem{Mark}
  K.~Ishiwata and M.~B.~Wise,
  ``Higgs Properties and Fourth Generation Leptons,''
  Phys.\ Rev.\ D {\bf 84} (2011) 055025
  [arXiv:1107.1490 [hep-ph]].
  
  
  
\end{thebibliography}
\end{document}